\documentclass[a4paper, keeplastbox
              ]{jacow}
\makeatletter%
	\ifboolexpr{bool{xetex}}
	 {\renewcommand{\Gin@extensions}{.pdf,%
	                    .png,.jpg,.bmp,.pict,.tif,.psd,.mac,.sga,.tga,.gif,%
	                    .eps,.ps,%
	                    }}{}
\makeatother

\ifboolexpr{bool{xetex} or bool{luatex}} 
 {}                                      
 {\usepackage[utf8]{inputenc}}           

\usepackage[USenglish]{babel}			 

\usepackage[final]{pdfpages}
\usepackage{multirow}
\usepackage{ragged2e}

\ifboolexpr{bool{jacowbiblatex}}%
 {%
  \addbibresource{jacow-test.bib}
  \addbibresource{biblatex-examples.bib}
 }{}
\begin{document}

\title{Recent Developments in DEMIRCI, the RFQ Design Software}

\author{
  E. Celebi\textsuperscript{1}\thanks{emre.celebi@cern.ch}, Istanbul Bilgi University, Faculty of Engineering and Natural Sciences, Istanbul, Turkey\\
  O. Cakir, 
  G. Turemen\textsuperscript{2}, 
 B. Yasatekin\textsuperscript{2}, Ankara University, Department of Physics, Ankara, Turkey\\
 G. Unel, University of California at Irvine, Department of Physics
 and Astronomy, Irvine, USA\\
 \textsuperscript{1} also at Bogazici University, Department
 of Physics, Istanbul, Turkey\\
 \textsuperscript{2} also at SANAEM, Ankara, Turkey
}
	
\maketitle

\begin{abstract}
The RFQ design tool DEMIRCI aims to provide fast and accurate simulation of a light ion accelerating cavity and of the ion beam in it. It is a modern tool with a graphical user interface leading to a "point and click" method to help the designer. This article summarizes the recent developments of DEMIRCI software such as the addition of beam dynamics and 8-term potential coefficient calculations. Its results are compared to other software available on the market, to show the  attained compatibility level. Finally the future prospects are discussed.

\end{abstract}

\section{Introduction}

Since their proposal by Kapchinsky and Tepilyakov (K-T) \cite{TepilKap},
the radio frequency quadrupoles (RFQ) are being used to bunch, focus
and accelerate ion-beams to energy levels high enough to be used at
the drift tube linacs (DTL) with various degrees of success. The successful
operation of an RFQ, requires a rapid and realistic design and simulation
of the cavity and the ion beam within. An accurate and easy design process
has proven to be a challenge since the specifications dictated by
research community and industry become ever more demanding. The complexities
in low energies makes it harder to design and to make simple predictions
based on that design. Using multiple computer intensive simulations at
each iteration and manually keeping track of the various design tools
used together, is error prone and time consuming. There is a limited
number of such computer programs, PARMTEQ and LIDOS being the leading
examples \cite{Parmteq,Lidos}. DEMIRCI was born out of the need for
a simple-to-use, yet a tool with multiple capabilities. Its first
version \cite{demirint,demirpaper}, allowed a very basic design using only
two term potential but provided the graphical means for helping the
inexperienced user. Since than the variable set that can be plotted
has been enlarged, it was made multi-lingual and available in windows
operating system. The details of these enhancements are discussed
elsewhere \cite{Dem-recent}. This note will focus on much more recent
developments such as eight term potential multipole coefficients calculation,
beam dynamics simulations with eight or two terms and RFQ acceptance
calculations and plots.

\section{Recent Developments}

The recent work has been focusing on the calculation of the eight
term (8T) potential multipole coefficients in a fast and accurate
way, on using these coefficients to track the macro-particles in the
RFQ, and finally on using this mechanism to both visualize the beam
behavior and to calculate the RFQ acceptance and mismatch factor
for a given low energy beam. 

\subsection{Eight Term Potential }

Designing an RFQ, consisting of about hundred cells, necessitates
correct determination of the time independent part of the electrical
K-T potential for each cell \textit{\emph{given \cite{TepilKap}by:}} 

\begin{eqnarray}
U(r,\theta,z) & = & \frac{V}{2}[\sum_{m=1}^{\infty}A_{0m}(\frac{r}{r_{0}})^{2m}\cos(2m\theta)\label{eq:generic-potential}\\
 & + & \sum_{m=0}^{\infty}\sum_{n=1}^{\infty}A_{nm}I_{2m}(nkr)\cos(2m\theta)\cos(nkz)]\nonumber 
\end{eqnarray}
where $r$ and $\theta$ are spherical coordinates for which $z$
represents the beam direction, $V$ is the inter-vane voltage, $k$
is the wave parameter given by $k\equiv2\pi/\lambda\beta$, with $\lambda$
being the RF wavelength and $\beta$ being the speed of the ion relative
to the speed of light. Also, $r_{0}$ is the mean aperture of the
vanes, $I_{2m}$ is the modified Bessel function of order $2m$ and
the $A_{nm}$ are the multipole coefficients whose values, depending
on the vane geometry, should be obtained. The heritage from the very
first RFQ studies in LANL can still be found in the industry standard
software PARMTEQ's \cite{Parmteq} method to determine the multipole
coefficients of the first 8 terms. It is based on data from pre-prepared
tables containing image charge calculations using the integral method.
Another approach reported in the literature uses a 3 dimensional differential
finite element method to obtain the potential distribution across
the RFQ length and then does a least squares fit to obtain the multipole
coefficients \cite{FEM-first}. Since this method is reported to be
as accurate as the image charge method and much faster from computation
point of view, an independent implementation of this approach is realized
in DEMIRCI.

\begin{figure}
   \centering
   \includegraphics*[width=174pt]{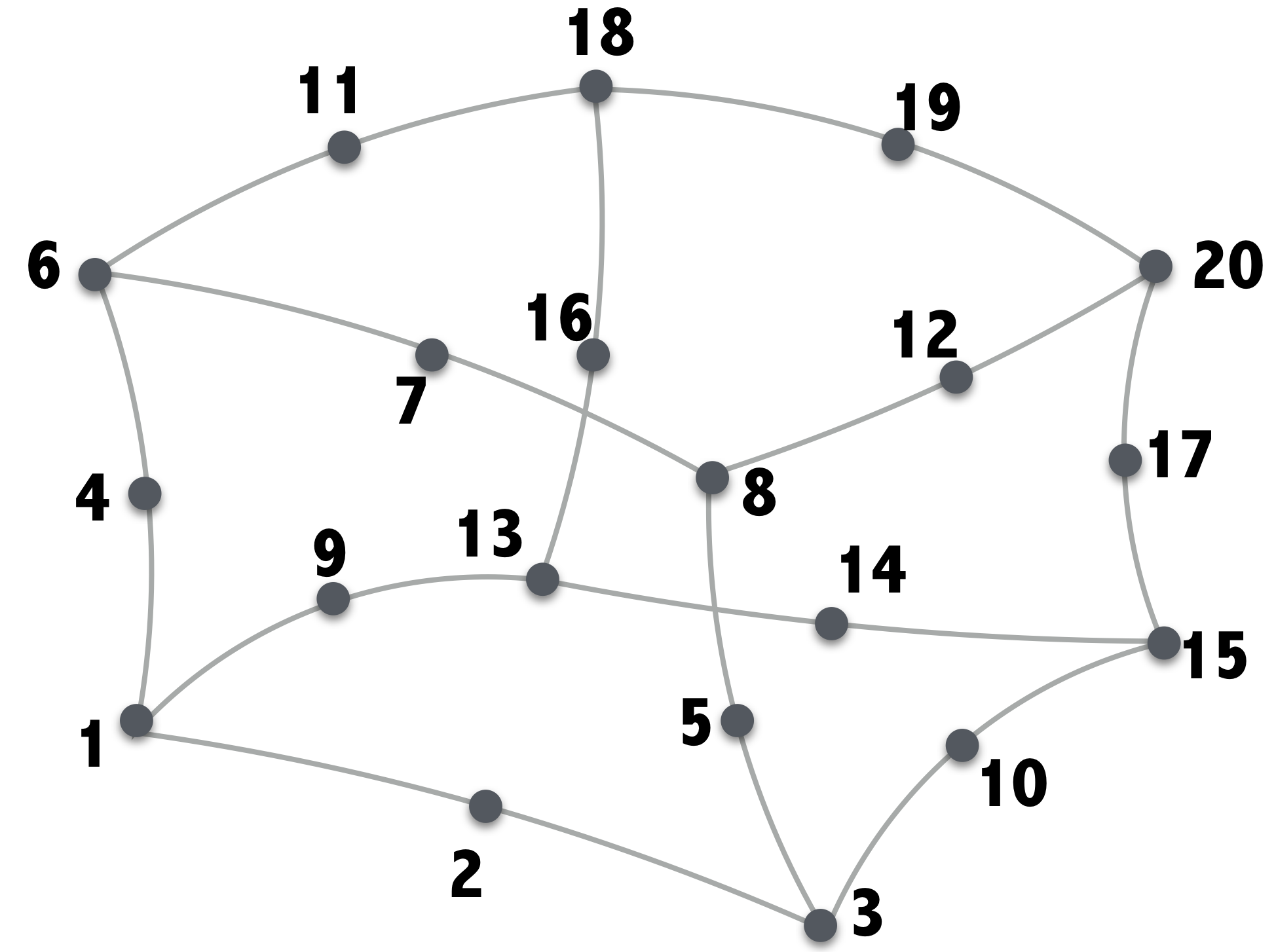}
   \caption{a typical 20 node isometric brick.}
   \label{20node}
\end{figure}

The procedure for finding the 8T coefficients starts with the division of the RFQ to be simulated into a number 20 node bricks as shown in Figure \ref{20node}. Although the number of segmentations in 3D is user selectable, a good compromise between speed and accuracy is encoded as the default value set: 5x5x6 in $x$, $y$ and $z$ coordinates. Once the meshing is complete, the general stiffness matrix is prepared from individual segments.  Finite element method (FEM) is used to formulate the Poisson equation to find the electrostatic potential for each cell. This equation is solved numerically using the conjugate gradient technique for each node and for each cell. The solutions for the nodes within the minimum bore radius are then fitted to the 8T potential function with the least squares method to find the first eight KT potential coefficients at each cell.

\begin{figure}
   \centering
   \includegraphics*[width=174pt]{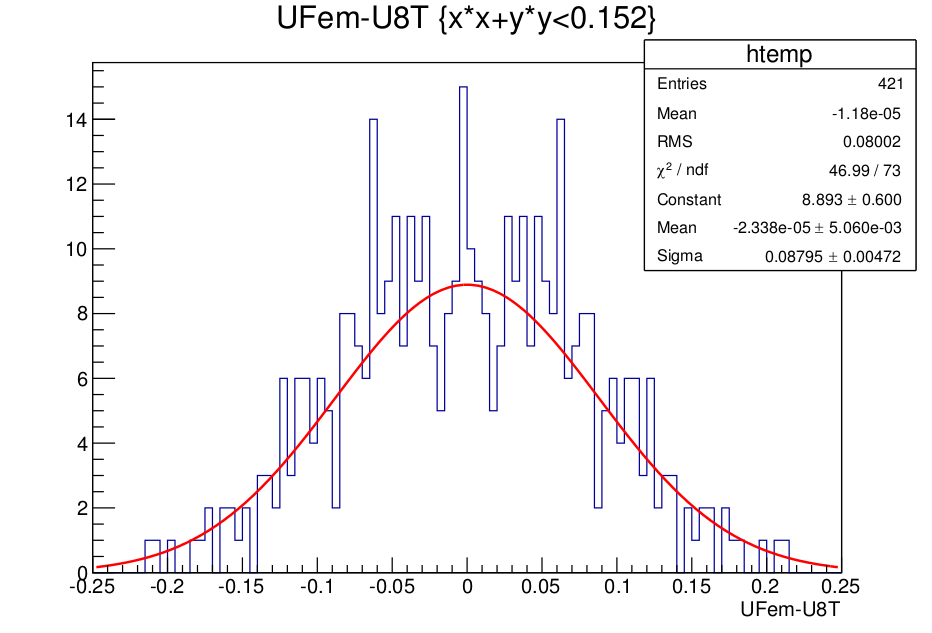}
   \caption{20th cell FEM and 8 term potential difference.}
   \label{20hucregaus}
\end{figure}

The validity of the method and of the results are checked using two different approaches. Firstly, a similar RFQ 8T coefficient calculation report has been found and the same calculations are repeated in DEMIRCI. The encouraging results of that study are reported elsewhere \cite{demirci2}. As an additional check, FEM solutions on all nodes are compared to fit results  for a random cell. The histogram of the difference between the two are shown in Figure \ref{20hucregaus} together with a gaussian fit. The distribution shows that the error originating from the fitting procedure is about 1/1000. Even if in some long cells, the 8T potential is not adequate for defining the equipotential surface, the error is smoothed out by the fitting procedure and it's order of magnitude is expected to remain at this level.

\begin{figure}
   \centering
   \includegraphics*[width=174pt]{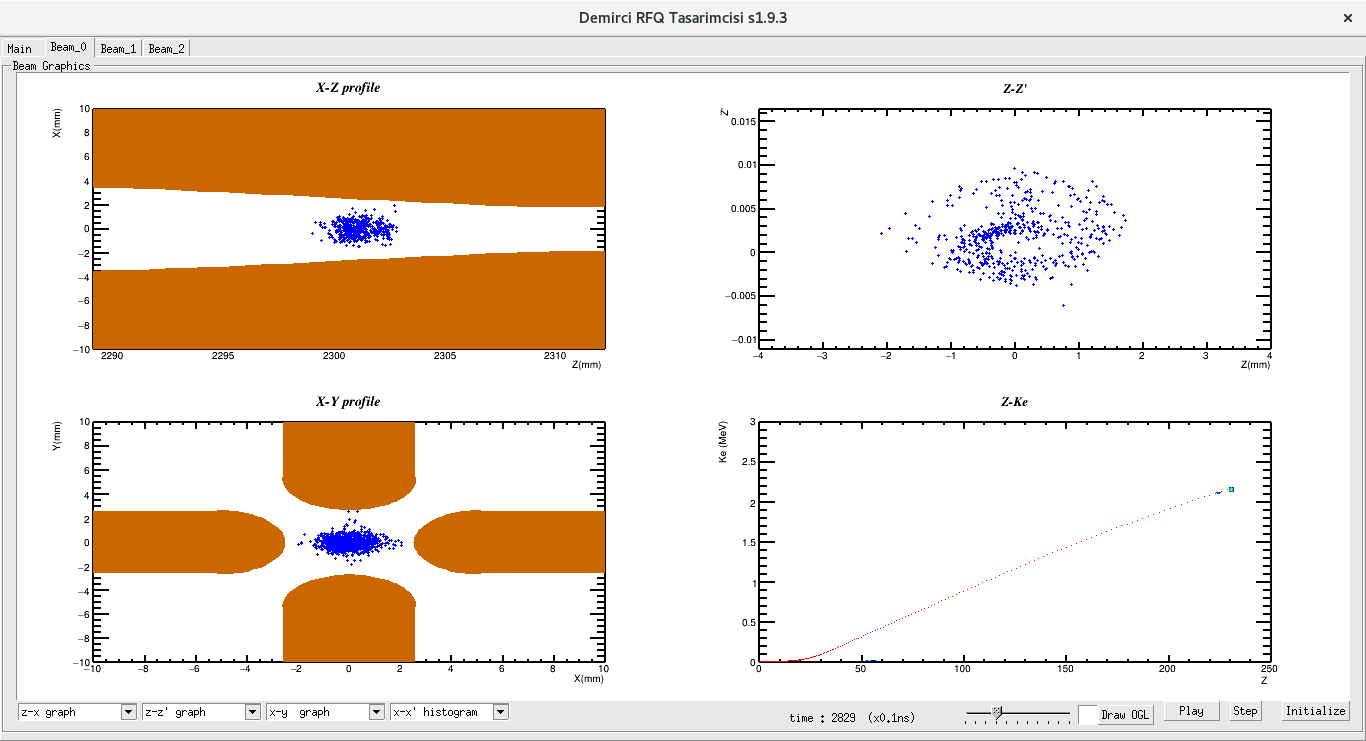}
   \caption{DEMIRCI beam dynamics tab during a simulation.   \label{bd-tab}}
\end{figure}

\begin{figure}
   \vspace*{-.5\baselineskip}
   \centering
   \includegraphics*[width=174pt]{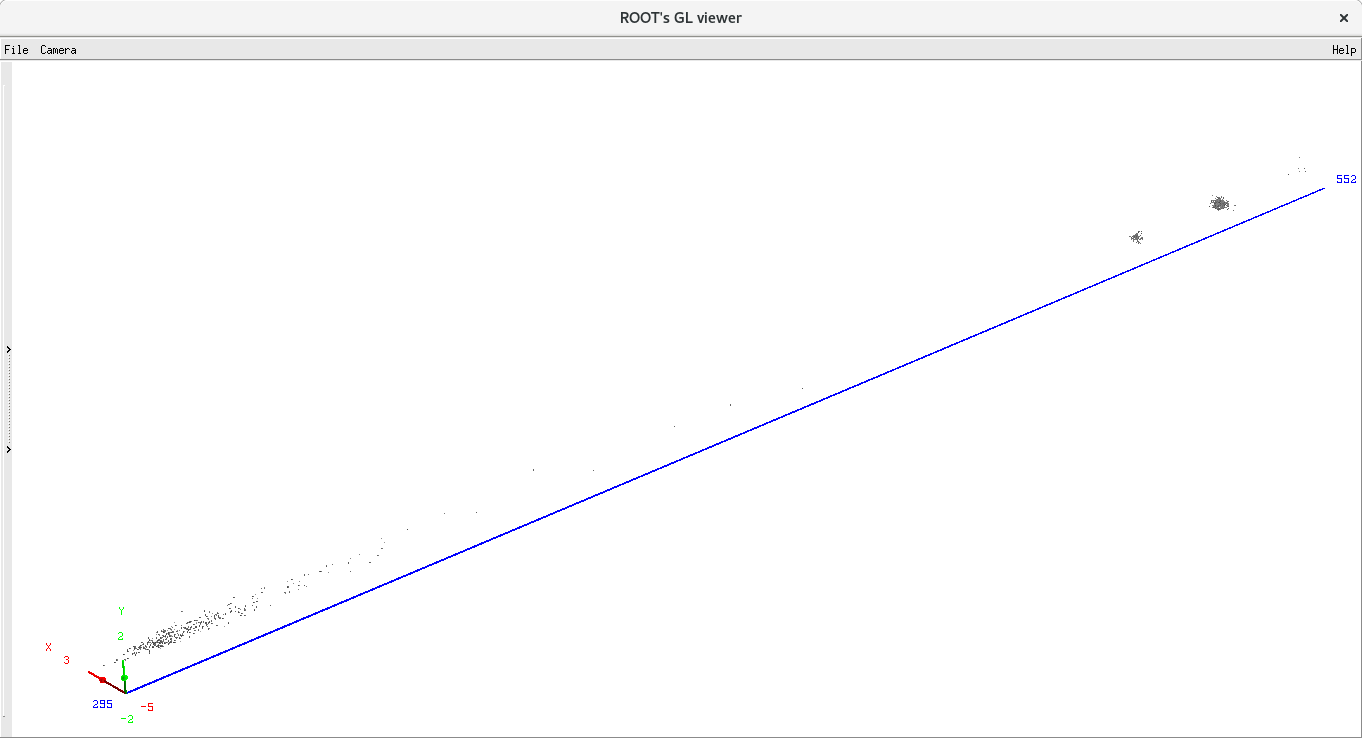}
   \caption{3D plot showing bunched and accelerated particles (synchronous particle is located around 520 mm).
   \label{dem3D}}
\end{figure}

\begin{figure}[h]
  \vspace*{-.5\baselineskip}
    \centering
    \includegraphics*[width=174pt]{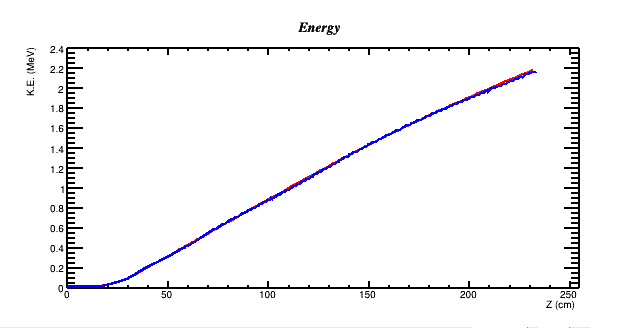}
    \caption{Expected kinetic energy (red) and synchronous particle energy calculated with 8 Term potential(blue)}
    \label{enerji}
 \end{figure}

Once all the 8T potential coefficients are calculated, the Beam Dynamics tabs of the DEMIRCI graphical user interface, shown in Figure\ref{bd-tab}, become available to the designer. The designer has three such tabs with four user selectable beam dynamics related plots are available. For the beam dynamics calculations a constant time step (set as 0.1ns, not adjustable by the user) based approach is used for finding all relevant EM fields, particle positions and velocities during the simulation. Synchronous particle energy calculated with these fields closely follows the predicted kinetic energy for our test design as shown in the Figure \ref{enerji}. Another visualization facility available to the designer is the 3D beam view shown in Figure \ref{dem3D}. In this particular example, where the synchronous particle is at 520mm from the origin, the bunched and accelerated particles are shown along the $z$ axis.

\section{Comparisons with similar software}

\begin{figure}
   \centering
   \includegraphics*[width=174pt]{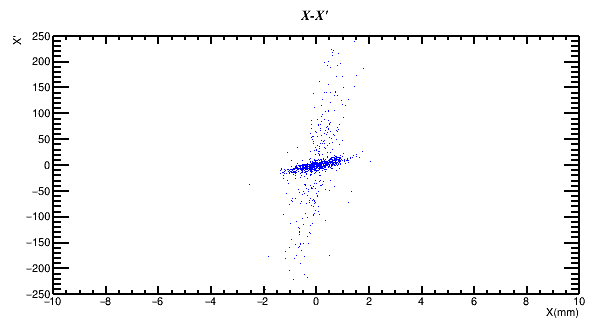}\\
   \includegraphics*[width=174pt]{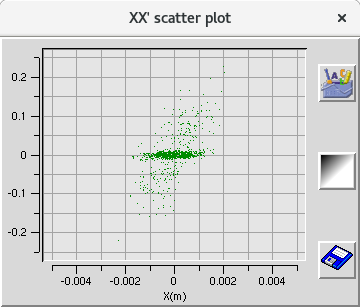}
      \caption{Protons in x-x' around z=2300 mm, top DEMIRCI, bottom TOUTATIS, using the same design.}
   \label{xxp_dem_tut}
\end{figure}

\begin{figure}
   \centering
   \includegraphics*[width=174pt]{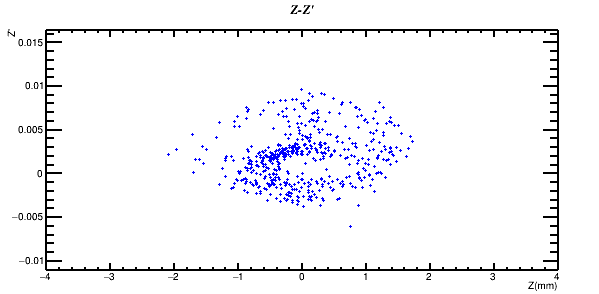}\\
      \includegraphics*[width=174pt]{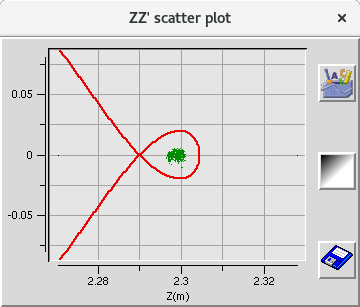}
   \caption{ Result for z-z' around 2300 mm top DEMIRCI, bottom TOUTATIS.}
   \label{zzp_dem_tut}
\end{figure}

To ensure compatible results with already available software, the same RFQ design has been fed into both DEMIRCI and TOUTATIS\cite{toutatis}. Figure \ref{xxp_dem_tut} shows the $x' x$ phase space at around $z$=2300mm as obtained from both DEMIRCI and TOUTATIS. The $x'$ is defined as\cite{linacbook}:

\begin{equation}
x' = (x_f-x_i)/(z_f-z_i)
\end{equation}

where the subscript $f$ ($i$) stands for the next (current) position of a particle. A similar plot set, shown in Figure\ref{zzp_dem_tut} is also made for the  $z' z$ phase space where $z'$ is defined using the velocity of the synchronious particle ( $Vz_{sync}$) along the $z$ direction:

\begin{equation}
z'=(Vz_i - Vz_{sync}) / Vz_{sync} \quad .
\end{equation}

One can thus see the relation between the $z$ position of the particles and their velocities showing the details of the bunching process. Both plots have DEMIRCI results at the top and TOUTATIS results at the bottom. Although it is hard to read the values from the TOUTATIS plots, the order of magnitude of the phase spaces coincide and both programs show similar bunching structures.

\section{Outlook}
DEMIRCI is being developed continuously to match the requirements of both national and international RFQ designer communities. Its graphically oriented easy to use user interface together with its fast and accurate calculations could make it the tool of choice for designers. To improve its simulation capabilities at high current values, both image and space charge simulations will be added in the next versions. Additionally a new project for producing a short RFQ to compare its EM fields to DEMIRCI predictions is submitted for funding.

\section{Acknowledgment}

This project has been supported by TUBITAK with project number 114F106.

\end{document}